\documentstyle[12pt,doublespace]{article}
\begin{document}
\newcommand{\scr}{\sin^2 \hat{\theta}_W (M_Z)}
 \newcommand{\sef}{\sin^2 \theta_{eff}^{lept}}
\newcommand{\smallms}{{\scriptscriptstyle \rm MS}}
\newcommand{\smallmsbar}{\overline{\smallms}}
\newcommand{\msbar}{\rm{\overline{MS}}}
\newcommand{\be}{\begin{eqnarray}}
\newcommand{\en}{\end{eqnarray}}
\newcommand{\mms}{\hat{m_t}(m_t)}
\newcommand{\mmss}{\hat{m_t}^2(m_t)}
\newcommand{\mpo}{m_t}
\newcommand{\scar}{\hat{s}}
\newcommand{\cc}{\hat{c}}
\newcommand{\as}{\alpha_s}
\newcommand{\amzc}{\hat{\alpha}(M_Z)}
\newcommand{\arun}{\alpha_{\rm run}}
\newcommand{\alc}{\hat{\alpha}}
\newcommand{\nn}{\noindent}
\newcommand{\ew}{electroweak\ }
\newcommand{\ewc}{electroweak corrections\ }
\newcommand{\dres}{(\Delta r)_{\rm res}}
\newcommand{\dr}{\Delta \rho}
\newcommand{\drf}{(\Delta \rho)_f}
\newcommand{\PL}{Phys. Lett.\ }
\newcommand{\NP}{Nucl. Phys.\ }
\newcommand{\PR}{Phys. Rev.\ }
\begin{titlepage}
\begin{flushright}
 NYU--TH--94/11/02 \\
 hep-ph/9411363\\
 November 1994 \\
\end{flushright}
\vspace*{2cm}

\centerline{\large{\bf On the QCD corrections
 to $\Delta\rho^\dagger$ }}

\vspace*{1.6cm}

\centerline{\sc Alberto Sirlin.}

\vspace*{.9cm}

\centerline{ Department of Physics, New York University, 4 Washington
Place,}
\centerline{ New York, NY 10003, USA.}

\vspace*{1.2cm}
\begin{center}
\parbox{14.6cm}
{\begin{center} ABSTRACT \end{center}
\vspace*{0.2cm}
We discuss some recent developments in the evaluation of the QCD corrections
to $\dr$, their interpretation,
an estimate of the theoretical error, and its effect on \ew physics.
}

\end{center}
\vspace*{4cm}
$^\dagger$ To appear in ``Reports of the Working Group on Precision
Calculations for the Z-resonance''.

\end{titlepage}
\newpage
\nn
The $\rho$ parameter is conventionally defined as the ratio of the effective
neutral and charged-current couplings at $q^2=0$. Although in the Standard
Model (SM) the associated radiative correction $\Delta \rho\equiv1-\rho^{-1}$
is process-dependent,
\footnote{See, for example, the second paper in Ref.[1].}
its fermionic component, $\drf$, is universal. We recall
that its contribution, $\rho_f=[1-\drf]^{-1}$,  is frequently separated out,
as an overall renormalization factor, when neutral
currents amplitudes are expressed in terms of $G_\mu$, even when
one considers amplitudes with $q^2\sim M_Z^2$.
Furthermore, $\drf$ is related to the leading asymptotic
corrections,
for large $m_t$, of the basic radiative corrections
$\Delta r$, $\Delta\hat{r}$, and $\Delta\hat{\rho}$ \cite{1}\cite{2}.
For these reasons, it is highly desirable to evaluate $\drf $
as accurately as possible, and to obtain an estimate of the theoretical
error  due to the unknown higher order corrections.
In particular, the study of the QCD corrections to $\drf$ has
recently been the subject of considerable attention.

Neglecting higher order \ew effects $\sim (\alpha/\pi s^2 )^2
(m_t^2/M_W^2)^2$, but retaining QCD corrections, $\drf$
can be written as
\be
(\dr)_f= \frac{3G_\mu m_t^2}{8\sqrt{2} \pi^2}\  [1 + \delta_{QCD}],
\en
where  $\mpo$ is the pole mass, the first factor is the one-loop result
\cite{3}, and  $\delta_{QCD}$ represents the relevant QCD correction.
 The $O(\as)$ contribution was obtained some time ago by Djouadi and
Verzegnassi \cite{4}
\be
\delta_{QCD}= -\frac{2}{9}(\pi^2+3) a(\mpo)+ ...=
- 2.860 a(\mpo)+ ...\label{dj}
\en
where $a(\mu)\equiv \as(\mu)/\pi$.
Attempts to go beyond this result have been carried out via two
different methods:
the dispersive approach, pioneered  by Kniehl, Kuhn, and Stuart \cite{5},
 and the direct examination of the relevant Feynman amplitudes at
$q^2=0$. The dispersive approach and its comparison with other
calculations is reviewed in the contribution of B.A. Kniehl to these
Proceedings and in Ref.\cite{knie}. The present contribution only
discusses recent developments in
the Feynman diagram approach, their interpretation, and an estimate
of the theoretical error.

Very recently,  Avdeev, Fleischer, Mikhailov, and Tarasov \cite{6}
carried out
a complete three-loop calculation of $O(\alpha\as^2)$. Their result,
obtained in the limit $m_b\to 0$, can be expressed to good accuracy as
\be
\delta_{QCD}= c_1 a(\mpo) + c_2 a^2(\mpo) + ...,
\label{avd}\en
where $c_1=-2.860...$ is the Djouadi-Verzegnassi result, and
\be
c_2=-21.271 + 1.786 N_f = - 10.55...\label{c2}
\en
In Eq.(\ref{c2}) $N_f=6$ is the total number of quarks
contributing via the vacuum polarization loops (5 massless quarks and
the top).
The contribution of the top quark is very different from the that of the
massless quarks and has been split in two parts: one, corresponding
to a ``massless top'' (hence the factor $N_f=6$) and the remainder,
which is included in the first term.
As pointed out in Ref.\cite{6},
it is also interesting to note that about 40\% of $c_2$ arises
from the ``anomaly-type diagrams'',
where the $\bar t \gamma^\mu \gamma_5 t$ currents are attached to triangle
diagrams linked by two virtual quarks \cite{7}.

In order to discuss the evaluation of $\as(\mu)$ in these expressions,
we recall that this parameter conventionally evolves with five active flavors
for $m_b<\mu<\mpo$, and with six for $\mu>\mpo$. At $\mu=\mpo$,
$\as(\mu)$ is continuous through $O(\as^2)$. There is a very small
discontinuity, $-(25/72) a^3(\mpo)$\cite{8}, but that occurs beyond the order
of current $\drf$ calculations and, moreover, is negligibly small.
For the purpose of our discussion, we can therefore treat $\as(\mu)$
as being continuous at $\mu=\mpo$. From these observations it follows
that $\as(\mpo)$ in Eq.(\ref{dj}) should be evaluated
conventionally, evolving $\as(\mu)$ from $\as(M_Z)$ with a 3-loop $\beta$
function and five active flavors.
As $\drf$ and $\mpo$ are physical observables, they are $\mu$-independent.
Therefore the same is true of $\delta_{QCD}$.
However, because Eq.(\ref{avd}) is truncated in $O(\as^2)$, its evaluation
depends somewhat on the chosen scale.
The $\mu$-dependence of Eq.(\ref{avd}) can be studied by using the simple
relations
\be
\as(\mpo)&=& \as^{(5)}(\mu)\left[
1- \beta_1^{(5)} \frac{\as^{(5)}(\mu)}{\pi}\ln(\frac{\mu}{\mpo})
\right] + ...\nonumber\\
&=&\as^{(6)}(\mu)\left[
1- \beta_1^{(6)} \frac{\as^{(6)}(\mu)}{\pi}\ln(\frac{\mu}{\mpo})
\right] + ...,
\label{runn}
\en
where $\beta_1^{(n_f)}= - \frac{1}{2}(11- \frac{2n_f}{3})$,
the ellipsis stand for higher order terms, and the superscript $n_f$ in
$\as^{
(n_f)}$ and $\beta_1^{(n_f)}$ represents the number of active flavors
in the evaluation of $\as(\mu)$. Eq.(5) follows from the fact that
$\as^{(5)}(\mu) $ and $\as^{(6)}(\mu)$ satisfy the RG equations with $n_f=5$
and $n_f=6$, respectively, and from the above mentioned continuity at
$\mu= \mpo$. Some authors employ $\as^{(6)}(\mu)$ for $\mu<\mpo$ as well
as for $\mu>\mpo$. That choice, although  theoretically acceptable,
is inconvenient because the experimental value of $\as(M_Z)$
should be identified with $\as^{(5)}(M_Z)$ rather than $\as^{(6)}(M_Z)$.
Therefore, it is natural to use the expression involving
$\as^{(5)}(\mu)$ in Eq.(\ref{runn}) when $m_b<\mu<\mpo$, and that in
terms of $\as^{(6)}(\mu)$ for $\mu>\mpo$. For clarity,
the following related observations are relevant:
i) The suggestion has been made by some theorists that,
in replacing $\as(\mpo)$ by the r.h.s. of Eq.(\ref{runn}) with
$n_f=5$ and
$\mu<\mpo$, one should also change $N_f$ from 6 to 5 in
the $\mu$-independent coefficient $c_2$.
We see, however, that such procedure is theoretically
inconsistent. While the shift from $n_f=6 $ to $n_f=5$ in Eq.(\ref{runn})
is re-absorbed in a redefinition of $\as(\mu)$ in a $\mu$-independent
manner, a change of $N_f$ in Eq.(\ref{c2}) for $\mu<\mpo$
makes $\delta_{QCD}$ $\mu$-dependent in $O(\as^2)$.
This contradicts the basic requirement that $\delta_{QCD}$
should be $\mu$-independent through the order of the calculation.
Or, putting this in a slightly different language, $\mpo$ is
$\mu$-independent and, unlike $\hat{m_t}(\mu)$, it cannot
be adjusted to absorb a $\mu$-dependent contribution from
$\delta_{QCD}$.
ii)  Some theorists evaluate $\as(m_t)$ by evolving
$\as^{(6)}(\mu)$ from $\as(M_Z)$. As $\as(M_Z)$ should be identified with
$\as^{(5)}(M_Z)$, we see from Eq.(\ref{runn}) that such approach introduces
a small but unnecessary error.

In this report we follow the above observations: in the range
$m_b<\mu<\mpo$, $\as(\mu) $ is evaluated by evolving $\as(\mu)$ from
$\as(M_Z)$
with a three-loop $\beta$ function\footnote{See, for example, Eq.(A.2) of
Ref.\cite{bra}.}
and five active flavors.
For definiteness, we take $\as(M_Z)=0.118$ and $M_Z=91.19$GeV,
and adjust $\Lambda_{\msbar}^{(5)}$ accordingly.
In order to discuss the $\mu$-dependence of the truncated series
or to optimize the perturbative expansions, we employ Eq.(\ref{runn}),
choosing the five-flavors expression in the domain $m_b<\mu<\mpo$.
As an illustration, for $\mpo=200$GeV, Eq.(\ref{avd}) leads to $\delta_{QCD}=
-0.0961-0.0119= -0.1080$, so that the $O(\as^2)$ term implies an enhancement
$0.0119/0.0961= 12.4\% $ of the leading QCD result.
If $\as(\mpo)$ in Eq.(\ref{avd}) is expressed in terms of
$\as(\mu)$ and the resulting series truncated in $O(\as^2)$,
for $0.1\le\mu/\mpo\le 1$ we find a variation $\le5.8\times10^{-3}$,
a bound that amounts to $5.4\% $ of the total QCD correction or $49\%$
of the $O(\as^2)$ term. It is difficult, however, to make a precise statement
of the error based on these considerations because the $\mu$-dependence of the
 truncated series depends on the chosen interval.
For example, in this case the variation is significantly smaller in the
interval $0.15<\mu/\mpo <1$ and significantly larger in the range
$0.075<\mu/\mpo<1$.
Furthermore, the mildness of the $\mu$-dependence of the first N terms of a
QCD expansion over an interval is consistent with but does not imply the
smallness of the higher order terms. The point is that, for example, the
$\mu$-dependence induced by a significantly large $O(\as^{N+1})$ term
is of $O(\as^{N+2})$ and may be cancelled by  sizable contributions
of that and higher orders.
 The application of the
Brodsky-Lepage-Mackenzie (BLM)\cite{9}, Principle of Minimal Sensitivity
(PMS)\cite{10},
and Fastest Apparent Convergence (FAC)\cite{11} methods to optimize
the scale in Eq.(\ref{avd}) is discussed later on.

On the other hand, we note that the expansion in Eq.(\ref{avd})
involves rather large and increasing coefficients, a feature that frequently
indicates significant higher order terms.
Furthermore,  arguments advanced in Ref.\cite{12} suggest that there are
  at least two scales in $\delta_{QCD}$: one, of $O(\mpo)$, associated with
the intrinsic corrections to the \ew amplitude, and another one, much smaller,
 related to  contributions to the pole mass $\mpo$
involving small gluon momenta. It is therefore a good idea
 to find alternative expressions for
$\delta_{QCD}$ that separate the two scales, and at the same time involve
terms of $O(\as^2)$ with  coefficients of $O(1)$ rather than $O(10)$.
The advantage of this  strategy is explained later on, when we discuss
the theoretical error.
 A simple way of implementing
this idea has been outlined in Ref.\cite{13}:
one expresses first $(\dr)_f$ in terms of $\mms$, the running $\msbar$
mass evaluated
at the pole mass, and then relates $\mms$ to $\mpo$ by optimizing the
expansion for $\mpo/\mms$, which is known through $O(\as^2)$.
Recently these arguments have been considerably refined \cite{14}
using the new results of Ref.\cite{6}.

Calling $\mu_t$ the solution of
$\hat{m_t}(\mu)= \mu$  and using Eq.(19) of Ref.\cite{6}, we have
\be
\drf =  \frac{3G_\mu \mu_t^2}{8\sqrt{2} \pi^2}\  [1 + \delta_{QCD}^{\msbar}],
\label{ms1}
\en

\be
\delta_{QCD}^{\msbar}= - 0.19325 \ a(\mu_t) +0.07111 a^2(\mu_t).
\label{ms}
\en
 We note that the convergence pattern of Eq.(\ref{ms}) is
very nice, with  very small leading and next-to-leading  coefficients.
 For this reason we will assume that the terms of $O(\as^3)$
 and higher are negligible in Eq.(\ref{ms1}) and evaluate
Eq.(\ref{ms}) with $\mu_t\to \mpo$
as this introduces only a small change of $O(\as^3)$.

We can now  express $\drf$ in terms of $\mms$, by  using  the
NLO expansion \cite{14}
\be
\frac{\mu_t}{\mms}= 1 + \frac{8}{3} a^2(\mpo)
 + [35.96 - 2.45 n_f]a^3(\mpo)
\label{five}
\en
where $n_f=5$ is the number of light flavors.
Defining $\Delta_{QCD}$ by
\be
\drf= \frac{3G_\mu \mmss}{8\sqrt{2}  \pi^2} \ [1 + \Delta_{QCD}],
\label{sei}
\en
we have
\be
1+\Delta_{QCD}= \left( \frac{\mu_t}{\mms}\right)^2 \left[
1+ \delta_{QCD}^{\msbar}\right].
\label{nine}
\en
Neglecting terms of $O(\as^3)$, we can combine Eqs.(\ref{ms1}-\ref{nine})
in a single expansion:
\be
\Delta_{QCD}= -0.19325\  a(\mpo)+C a^2(\mpo).
\label{sette}
\en
In Ref.\cite{13} Eq.(\ref{sette}) was proposed  before the results of
 Ref.\cite{6} became known, and it was argued, on the basis of convergence
assumptions, that $|C|\le6$.
\ From Eqs.\ref{ms}-\ref{five} we see that $C=16/3 + 0.071=5.40$, consistent
 with the arguments
of Ref.\cite{13}.\footnote{In Ref.\cite{13} C was also estimated
to be $\approx +3$ by optimization arguments,
but the more conservative value $C=0^{+6}_{-6}$ was employed
in the final analysis.}
On the other hand, with $C=5.40$ the two terms in Eq.(\ref{sette})
nearly cancel and the error estimate is unnecessarily large.
It is clearly better to insert in Eq.(\ref{nine}) the separate
expansions in Eqs.(\ref{ms})and (\ref{five}),
as the magnitude of the last terms in these expressions is significantly
smaller than in Eq.(\ref{sette}). In the case of Eq.(\ref{five}) we retain the
relatively large $O(a^3)$ contribution in order to control the scale
of the leading term. As its coefficient, 23.71,
is rather large, we apply the BLM
optimization procedure \cite{9} to Eq.(\ref{five}), and obtain
\be
\frac{\mu_t}{\mms}= 1
 + \frac{8}{3} a^2(0.252\mpo)
- 4.47 a^3(0.252\mpo) \ \ \ (BLM).
\label{otto}
\en
As an illustration, for $\mpo=200$GeV,
 Eqs.(\ref{five}) and (\ref{otto}) give $\mu_t/\mms=
1.00391$ and 1.00423, respectively.
The optimization of Eq.(\ref{five}) using the
PMS\cite{10} and the FAC\cite{11} methods
is very close numerically to  Eq.(\ref{otto})
 (the difference is $\le5\times 10^{-6}$).

Values for  $\delta_{QCD}^{\msbar}$ (Eq.(\ref{ms}) with $\mu_t\to\mpo$)
and $\Delta_{QCD}$ (evaluated via Eqs.(\ref{nine}), (\ref{ms})
 and (\ref{otto}))
are shown in Table 1. We see that these
are indeed small effects.
 In particular,
$\Delta_{QCD}=(2- 3)\times 10^{-3}$, depending on $\mpo$, a remarkably
small correction.
Comparing Eq.(1) and (\ref{sei}), we find
\be
1 +\delta_{QCD}=
\left(\frac{\mms}{\mpo}\right)^2 [1+ \Delta_{QCD}].
\label{ten}
\en
For $\mpo/\mms$ we have the well-known expansion \cite{15}
\be
\frac{\mpo}{\mms}=
1+\frac{4}{3} a(\mpo) +K \ a^2(\mpo),
\label{undici}
\en
In the limit of neglecting the masses of the first five flavors,
$K$ is given by \cite{8}\cite{16}
\be
K= 16.0065 - n_f 1.0414 + 0.1036,
\label{K}
\en
where the first term corresponds to the quenched approximation, and the
second and third are the vacuum polarization contributions
of the masseless quarks ($n_f=5$) and the top quark, respectively.
As expected, the latter is much smaller. Refs.\cite{6},\cite{15}
include the contribution of all flavors to the vacuum polarization and,
therefore, in those calculations
$\hat{m_t}(\mu)$ evolves with six active quarks in the $\beta$
and $\gamma$ coefficients. The same is true of $\hat{m_t}(\mu_t)= \mu_t$
in Eq.(\ref{ms1}) and $\mms$ in Eq.(\ref{undici}).
One can equivalently express these relations in
terms of $\hat{m_t}(\mu_t)$ and $\mms$
evaluated with five active flavors \cite{8}. Through terms of $O(\as^2)$ this
is simply done by ``decoupling'',
i.e. subtracting the small top contribution, 0.1036, from $K$ and
$2\times0.1036\approx0.21$ \ from the coefficient of $a^2$ \
in \ Eq.(\ref{ms}).\footnote{
It is worth emphasizing that even this small decoupling subtraction
should not be made in discussing the $\mu$-dependence of Eq.(1) because,
unlike $\hat{m_t}(\mu)$, $\mpo $ is $\mu$-independent.}
These changes are very small and, moreover, they are not necessary for
our purposes. As our aim is to combine Eqs.(\ref{ms1}) and (\ref{undici}),
both evaluated with six flavors, we have defined in the same way
$\mu_t$ and $\mms$ in Eq.(\ref{five}) et seq.,
without decoupling the small top quark contribution.

When Eq.(\ref{undici}) is inserted in Eq.(\ref{ten}), it induces a
contribution
$\sim -0.11 $ to $\delta_{QCD}$ while, as we saw before, $\Delta_{QCD}\sim
0.002$. Thus, we find the intriguing result that, when $\drf$ is expressed in
terms of $\mpo$, the correction is almost entirely contained in
$(\mms/\mpo)^2$, a pure QCD effect that can be studied in isolation  from
\ew physics. Once this is recognized, it becomes clear that the magnitude and
 error of $\delta_{QCD}$ are largely controlled by the value of
$(\mms/\mpo)^2$ and the accuracy within which it can be calculated. As $K$ is
quite large (10.90 for $n_f=5$), it is  natural
to apply the three well-known approaches \cite{9}-\cite{11}
to optimize the scale in Eq.(\ref{undici}). Given an $O(\as^2)$ expansion of
 the form $S=1+A a(\mpo)+ (B- C n_f) a^2(\mpo)$, where $A,\ B,$ and $C$ are
numbers, the application of these methods lead to the optimized expansions
\be
S=
1 +A \ a(\mu^*) +[B-\frac{33}{2}C]a^2(\mu^*) \ \ \ (BLM),
\label{blm}
\en
\be
S=
1 +A \ a(\mu^{**}) -\frac{A\beta_2}{2\beta_1}
a^2(\mu^{**}) \ \ \ (PMS),
\label{pms}
\en
\be
S=
1 +A\  a(\mu^{***})\ \ \ \ \  \ \ \ \ \ \ (FAC),
\label{fac}
\en
where
$\beta_2=-[51-19n_f/3]/4$,
 $\mu^*= e^{-3C/A} \mpo$, $\mu^{***}=e^{(B-Cn_f)/A\beta_1}\mpo$,
and $\mu^{**}= e^{(\beta_2/2\beta_1^2)}\mu^{***}$.
In deriving these expressions we have assumed that $\as$ evolves
with $n_f$ active flavors between $\mpo$ and the optimized scales.
 To obtain Eq.(\ref{pms}) we have employed the two-loop
RG differential operator to find out, to good accuracy, the stationary
point. This can be checked by inserting Eq.(\ref{runn}) in the r.h.s.
of Eq.(\ref{undici}) and evaluating numerically the series,
truncated in $O(\as^2)$, as a function of $\mu$.
Applying Eqs.(\ref{blm}-\ref{fac}) to Eq.(\ref{undici}),
\be
\frac{\mpo}{\mms}=
1 +\frac{4}{3} a(\mu^*) - 1.07 a^2(\mu^*) \ \ \ (BLM),
\label{12}
\en
\be
\frac{\mpo}{\mms}=
1 +\frac{4}{3} a(\mu^{**})- 0.84 a^2(\mu^{**})\ \ \ (PMS),
\label{13}
\en
\be
\frac{\mpo}{\mms}=
1 +\frac{4}{3} a(\mu^{***})\ \ \ \ \  \ \ \ \ \ \ (FAC),
\label{14}
\en
where $\mu^*= 0.0960 \mpo$,
$\mu^{**}= 0.1005\mpo$, $\mu^{***}=0.1185\mpo$.
In these expressions $\as$ is evaluated with $n_f=5$. As the optimization
scales are in the $m_b-\mpo$ range, this is consistent with the discussion
after Eq.(\ref{runn}).
For $\mpo=200$GeV, Eqs.(\ref{12}-\ref{14}) give $\mpo/\mms=1.06304$, 1.06303,
and 1.06297, respectively. For $\mpo=174$GeV, the corresponding values are
1.06477, 1.06477, 1.06470.
Finally, for $\mpo= 130$GeV, we have 1.06875, 1.06875, 1.06867.
Thus, the three approaches give expansions with similar scales, coefficients
 of $O(1)$, and remarkably close values.
In contrast, the expansion in Eq.(\ref{undici}), which involves
a large second order coefficient of  $O(10)$, gives
1.0571, 1.0584, 1.0614 for $\mpo=(200,174,130)$GeV, which are
 $(0.6- 0.7)\%$ smaller.
One must conclude that the coefficients
of the unknown terms of $O(a^3)$ and higher in Eq.(\ref{undici})
and or Eqs.(\ref{12}-\ref{14}) are large. For instance, if Eq.(\ref{12})
were exact, the coefficient of the
$a^3(\mpo)$  and $a^4(\mpo) $  terms in Eq.(\ref{undici})
would be $\approx$\ 104 and \ 1,041, respectively.\footnote{
For  recent applications of the BLM, PMS, and FAC approaches to estimate
 higher
order coefficients in other cases, see Ref.\cite{17},\cite{18}.}
 In the following we employ
the optimized expression for $\mpo/\mms$, which, for definiteness, we
identify with the BLM expansion (Eq.(\ref{12})). The advantage
of this procedure will be explained later on, when we discuss the theoretical
error.

Table 2 displays the values of $\delta_{QCD}$ obtained from Eq.(\ref{ten}),
using Eq.(\ref{12}) and our previous determination of $\Delta_{QCD}$
(Eq.(\ref{nine})), and compares them with those derived from Eqs.(\ref{avd},
 \ref{c2}).
In order to show the effect of the higher order contributions (H.O.C.),
 we also exhibit the
fractional enhancement of the total QCD correction over the conventional
$O(a)$ result (Eq.(\ref{dj})).
\ From Table 2 we see that, for 130GeV$\le\mpo\le$220GeV,
the evaluation of $|\delta_{QCD}|$ from  Eq.(\ref{ten})
leads to a $(18-20)\%$ enhancement over Eq.(\ref{dj}) and is
$(5.2- 6.6)\times 10^{-3}$ larger than the results from Eq.(\ref{avd}).
As a percentage of the total QCD correction this
 difference amounts to $\approx 5\%$, a reasonably small effect.
However, the last two columns in Table 2 show that the H.O.C.
 are  $\approx 45\%$ larger in the evaluation based
on Eq.(\ref{ten}).
Most of this divergence can be traced to different ways of treating the
dominant factor $(\mms/\mpo)^2 $ in Eq.(\ref{ten}). For example,
it is amusing to note that the expansion $(\mms/\mpo)^2 -1=
-(8/3) a(\mpo) - (2K- 16/3) a^2(\mpo)$, obtained from
Eq.(\ref{undici}), is quite close numerically to Eqs.(\ref{avd}, \ref{c2}).
For $\mpo=(200, 174, 130)$GeV it gives -0.1081,-0.1106, -0.1160, to
be compared with -0.1080, -0.1102, 0.1154 from Eqs.(\ref{avd}, \ref{c2})
(see Table 2). This conforms with our observation that the bulk of the QCD
correction to $\drf$ is contained in $(\mms/\mpo)^2 $, but at the same time
shows the main source of the difference between the two calculations.
In fact, in the evaluation of Eq.(\ref{ten}) we have employed
Eq.(\ref{12}), the optimized version of Eq.(\ref{undici}), rather than the
expansion mentioned above.

As $c_2$ in Eq.(\ref{c2}) is $O(10)$, in analogy with our discussion
of Eq.(\ref{undici}), we may directly attempt to optimize
Eqs.(\ref{avd},\ref{c2}). Applying Eqs.(\ref{12}-\ref{14}) to
Eqs.(\ref{avd},\ref{c2}) with $N_f=n_f+1=6$, we have
\be
\delta_{QCD}= - 2.860 \ a(0.154\mpo)+
9.99 \ a(0.154\mpo)\ \ \ (Eq.(\ref{avd});\ BML),
\label{17}
\en
\be
\delta_{QCD}= - 2.860 \ a(0.324\mpo)+
1.80 \ a^2(0.324\mpo)\ \ \ (Eq.(\ref{avd});\ PMS),
\label{18}
\en
\be
\delta_{QCD}= - 2.860 \ a(0.382\mpo)
\ \ \ \ \ \ \ \ (Eq.(\ref{avd});\ FAC).
\label{19}
\en
Once more, in Eqs.(\ref{17}-\ref{19}) $\as $ is evaluated with 5 active
flavors, in accordance with the discussion after Eq.(\ref{runn}).\footnote{
In Ref.\cite{6} the FAC scale is reported as being 0.348$\mpo$. The apparent
 difference with Eq.(\ref{19}) is due to the fact that $\as^{(6)}$ is employed
in that paper. In fact, both results are equivalent through $O(\as^2)$.}
For $\mpo=200$GeV, Eqs.(\ref{17}-\ref{19}) give 0.1084,
0.11045, 0.11038, respectively. The difference between
Eqs.(\ref{18},\ref{19}) and Eq.(\ref{ten}) is now only
$\approx 2.9\times10^{-3}$, which is $\approx 45\%$ smaller than that
between Eq.(\ref{avd}) and Eq.(\ref{ten}).
On the other hand, although the $a^2$ coefficient in Eq.(\ref{18})
is roughly of $O(1)$, that in Eq.(\ref{17}) is $O(10)$.
It has  been pointed out that in many NLO QCD calculations, it is a good
approximation to retain only the leading term, evaluated at the BLM scale
\cite{9},\cite{12}.
\ From Eq.(\ref{17}) we see that this is not the case
in the expansion of Eqs.(\ref{avd},\ref{c2}). Thus, in contrast to
Eqs.(\ref{12}-\ref{14}), when applied to Eqs.(\ref{avd},\ref{c2})
the three optimization  procedures do not  uniformly
lead to coefficients of $O(1)$ and similar scales.

A frequently used method to estimate the theoretical error
of QCD expansions is to consider the magnitude of the last included or known
term \cite{19}. By this criterion the dominant error in Eq.(\ref{ten})
is contained in the first factor. We employ Eq.(\ref{12}) as it is the
optimized expansion  with largest second order term. The corresponding error
is
\be
\delta\left(\frac{\mms}{\mpo}\right)^2
\approx\pm \frac{2\times 1.07 a^2(\mu^*)}{
(\mpo/\mms)^3} \approx \pm 1.77 a^2(\mu^*).
\label{25}
\en
For $\mpo=(200,174,130)$GeV, this amounts to $\pm(4.3,4.5,5.1)\times 10^{-3}$.
The calculation of the small correction $\Delta_{QCD}$ involves in turn two
uncertainties. One, associated with $(\mu_t/\mms)^2$, can be estimated
from Eq.(\ref{otto}) as
\be
\delta\left( (\mu_t/\mms)^2\right)\approx
\pm 8.9 a^3(0.252\mpo),
\en
and leads to $\pm (6,7,8)\times 10^{-4}$. The other, involving
$\delta^{\msbar}_{QCD}$, is given by
\be
\delta(\delta^{\msbar}_{QCD})\approx \pm 0.071 a^2(\mpo)
\en
and amounts to $\pm (8,8,9)\times 10^{-5}$. In Eq.(\ref{ten}) these two errors
are decreased by a factor $(\mms/\mpo)^2\approx 0.88$. Adding the three
uncertainties linearly we get an overall error estimate for Eq.(\ref{ten})
due to higher order corrections:
\be
\delta(\delta_{QCD})= \pm (4.9, 5.2, 5.9)\times 10^{-3} \ \ \  \ \ \
\left(Eq.(\ref{ten})\right),
\label{err}
\en
for $\mpo=200, 174, 130$GeV, respectively. It could be argued that the
smallness of the leading and NLO coefficients in Eq.(\ref{ms}) is fortuitous.
However, in order to lead to a contribution $\approx 2.5\times 10^{-3}$
for $\mpo=200$GeV, which would modify our error estimate by $50\% $, the
coefficient of $a^3(\mpo)$ should be $\approx 66,$ or 930 times the
coefficient
of $a^2(\mpo)$. If the same ``last term'' criterion were applied
to Eq.(\ref{avd}), the error estimate would be $\delta(\delta_{QCD})\approx
\pm c_2 a^2(\mpo)$, which amounts to $\pm (1.2, 1.2, 1.3)\times 10^{-2}$. This
is 2.2-2.4 times as large as in Eq.(\ref{ten}). The consideration of the
optimized expansions is very ambiguous in the case of
 Eqs.(\ref{avd}, \ref{c2}); while the PMS method (Eq.(\ref{18})) leads to a
 small error estimate, the BLM approach (Eq.(\ref{17})) gives a very large
 uncertainty $\approx\pm(2.0-2.4)\times 10^{-2}$.
This ambiguity may perhaps be related to the observation
that the expansion in Eq.(\ref{avd}) involves more than one scale.

We note that the two different evaluations of $\delta_{QCD}$, by means
of Eqs.(\ref{avd},\ref{c2}) and Eqs.(\ref{ten},\ref{nine},\ref{12}),
respectively, coincide through $O(\as^2(\mpo))$. The numerical difference
between the two means that at least in one of these two calculations there
are significant contributions of $O(a^3)$ and higher.
Although the present author prefers the latter approach, on the grounds that
it involves expansions with coefficients of $O(1)$ and leads by the ``last
term criterion'' to a significantly smaller error estimate,
it seems impossible to rigorously  decide at present which is
more accurate. This could be clarified, in principle, by evaluating the
$O(\as^3)$ terms in Eq.(\ref{avd}) and or Eq.(\ref{undici}).
Prospects for achieving this, however, appear to be quite remote \cite{16}.
On the other hand, we believe that Eq.(\ref{err}) is a reasonable estimate
of the theoretical error due to unknown higher order corrections.
In particular, we note that Eq.(\ref{err}) is also very close to the
 difference
between the two $\delta_{QCD}$ evaluations, which can also be used as an
estimate of the theoretical error \cite{14},
and to the scale variation of Eq.(\ref{avd}) in the interval
$0.1<\mu/\mpo<1$. Finally, one should remember
that there is at present an additional 5\% error in $\delta_{QCD}$
associated with the $\pm 0.006$ uncertainty in $\as(M_Z)$.
This leads to an additional contribution of $\pm(5.8,5.9, 6.2)\times10^{-3}$
to $\delta(\delta_{QCD})$, for $m_t=(200,174,130)$GeV.
In contrast to the theoretical error in Eq.(\ref{err}), it is likely that
this uncertainty can be decreased in the near future.

As pointed out in Ref.\cite{14}, the values for $\delta_{QCD}$ given
in Table 2 can accurately be represented by simple empirical formulae.
We find that the calculation based on Eq.(13) (third column of Table 2)
and the error estimate of Eq. (28) due to higher order corrections can be
conveniently expressed as
\be
\delta_{QCD}= -2.860 a(\xi m_t),
\en
\be
\xi=0.321^{+0.110}_{-0.073} \ \ \ \ \ (Eqs.(13,28)),
\en
while Eq.(3) (second column of Table 2) corresponds  \cite{14}
to $\xi=0.444$.
We emphasize that Eqs.(29,30) are not the result of a FAC optimization.
They are simply heuristic formulae that reproduce the values in the Tables
with errors of at most $\pm 1\times 10^{-4}$ for $\xi=0.321$ and $\xi=
0.431$, and at most $3\times 10^{-4}$ for $\xi=0.248$.
We also emphasize that in these expressions $\as$ is evaluated with five
active flavors in the manner explained before. We see that
Eq.(30) is somewhat smaller than the effective scale associated
with Eq.(3); however, both  evaluations are roughly consistent
within our theoretical error estimate.
By a numerical coincidence, the central value in Eq.(30) is very close to
the PMS scale in Eq.(23); however, Eqs.(29) and (23) differ
somewhat because of the presence of the $O(a^2)$ term in the latter.

In summary, in this report we have emphasized a number of results and
observations:
i) Working in the NLO approximation, we have pointed out that $\Delta_{QCD}$,
the QCD correction when $\drf$ is expressed in terms of $\mms$, is remarkably
small, $\approx (2-3)\times10^{-3}$.
This means that precision \ew physics essentially measures this parameter,
an important input for GUTs studies. The evaluation of $\Delta_{QCD}$
and $\delta_{QCD}^{\msbar}$ shows that $\mms$ and
$\mu_t\equiv\hat{m_t}(\mu_t)$
are very good expansion parameters as they absorb the bulk of the
QCD corrections \cite{20}.
Arguments for these desirable properties from effective field
theory have been given by Peris \cite{21}. If the smallness of $\Delta_{QCD}$
persists beyond  the NLO approximation, it also means that $\delta_{QCD}$,
the QCD correction when $\drf$ is expressed in terms of $\mpo$,
is almost entirely contained in $(\mms/\mpo)^2$, a pure QCD effect that can be
studied in isolation from \ew physics.
ii) We have also pointed out that when the BLM, PMS and FAC optimization
methods are applied to the expansion for $\mpo/\mms$, they lead to similar
scales, coefficients of $O(1)$ and remarkably close values.
Combining these optimized expansions with $\Delta_{QCD}$ one obtains an
evaluation of $\delta_{QCD}$ which shows an $(18-20)\%$ enhancement
over the two-loop calculation, depending on $\mpo$.
Using the magnitude of the last included terms as an estimate of the
theoretical error due to higher order corrections, this analysis
leads to $\delta(\delta_{QCD})\approx\pm (5-6)\times 10^{-3}$, depending
on $\mpo$. This estimate seems reasonable, as it is also close
to the difference between the $\delta_{QCD}$ evaluation
mentioned above and that obtained from the expansion
proposed by Avdeev et al..  If the theoretical error is combined quadratically
with that arising from $\delta\as$ one obtaines an overall uncertainty
in $\delta_{QCD}$ of $\pm(7.6,7.9,8.6)\times10^{-3}$
for $\mpo=(200,174,130)$GeV. For $\mpo=200$GeV and $M_H=300$GeV, this induces
errors $\pm 9.5\times10^{-5}$ in $\Delta\rho$,
$\pm3.1\times 10^{-4}$ in $\Delta r$, $\pm3.2\times 10^{-5}$ in the $\msbar$
parameter $\scr$, and $\pm 840$MeV and $\pm 5.5$MeV in the predicted masses
of $\mpo$ and $M_W$, respectively.
We also recall that these results scale approximately as $\mpo^2$,
so that they are actually smaller for $\mpo<200$GeV. Thus the effect
of these errors in \ew physics is rather mild. It should also be remembered
that the concept of pole mass has an intrinsic uncertainty $\sim\Lambda_{QCD}$
 and,  for the top quark, this may amount to 200-300MeV \cite{22}.
\vskip 1.3cm

\noindent{\bf{Aknowledgements}}
\vskip .4cm
The author is indebted to D. Bardin, D. Broadhurst, S. Brodsky, J. Fleischer,
P. Gambino, A. Kataev,
B. Kniehl,   W. Marciano, G. Passarino and L. Surguladze for  useful
discussions.
This work was supported in part by the National Science Foundation under
 Grant No. PHY--9313781.

\newpage
\begin{center}
$$
\begin{array}{|r|r|r|}\hline
  \mpo ({\rm GeV}) & \ 10^3 \delta_{QCD}^{\msbar}\
 & 10^3 \Delta_{QCD} \\ \hline
     130  & -6.80 & 2.92\\ \hline

   150 &-6.67& 2.59 \\ \hline

   174& -6.53& 2.28\\ \hline

   200& -6.41& 2.01 \\ \hline

 220& -6.33& 1.84 \\ \hline
   \end{array}
$$
\end{center}
\vskip 0.5cm
{\bf Table 1.} The corrections
$\delta_{QCD}^{\msbar}$ and $\Delta_{QCD}$. The first one is given by Eq.(7)
with $\mu_t\to\mpo$, while the second is obtained from Eq.(10),
with $\mu_t/\mms$ evaluated according to Eq.(12) ($\as(M_Z)=0.118$
 is employed).

\vskip 1.7cm
\begin{center}
$$
\begin{array}{|r|r|r|r|r|}\hline
  \mpo \ \  & \  \delta_{QCD}\  &  \ \delta_{QCD}\  & H.O.C.
& H.O.C. \\

  ({\rm GeV}) & ({\rm Eq.(3)}) & ({\rm Eq.(13)})& ({\rm Eq.(3)})\% &
({\rm Eq.(13)})\% \\ \hline

130  & -0.1154 & -0.1220 & 13.2 & 19.6\\ \hline

   150 &-0.1128& -0.1189 & 12.9 & 19.0\\ \hline

   174& -0.1102& -0.1160 & 12.6 & 18.5\\ \hline

   200& -0.1080& -0.1133 & 12.4 & 18.0\\ \hline

 220& -0.1064& - 0.1116 & 12.2 & 17.7\\ \hline
   \end{array}
$$
\end{center}
\vskip 0.5cm
{\bf Table 2.} Comparison of two determinations of $\delta_{QCD}$.
The second column is based on Eq.(3) [6]. The third column is based on
Eq.(13) with $\mpo/\mms$ obtained from Eq.(19) and $\Delta_{QCD}$ evaluated
according to Table 1. The fourth and fifth columns give the fractional
enhancement over the conventional $O(\as)$ result (Eq.(2)) due to the
inclusion of higher order contributions (H.O.C.).

\end{document}